\documentclass[aps,prl,twocolumn,showpacs,superscriptaddress]{revtex4}

\usepackage{amsmath}
\usepackage{graphicx}
\usepackage{dcolumn}
\usepackage{bm}

\begin{document}

\title{Quasiequilibrium supersolid phase of a two-dimensional
dipolar crystal}


\author{I.L.~Kurbakov}

\affiliation{Institute of Spectroscopy, 142190 Troitsk, Moscow region,
Russia}

\author{Yu.E.~Lozovik}

\affiliation{Institute of Spectroscopy, 142190 Troitsk, Moscow region,
Russia}

\author{G.E.~Astrakharchik}

\affiliation{Departament de F\'{\i}sica i Enginyeria Nuclear, Campus Nord B4-B5,
Universitat Polit\`ecnica de Catalunya, E-08034 Barcelona, Spain}

\author{J.~Boronat}

\affiliation{Departament de F\'{\i}sica i Enginyeria Nuclear, Campus Nord B4-B5,
Universitat Polit\`ecnica de Catalunya, E-08034 Barcelona, Spain}

\date{\today}

\begin{abstract}
We have studied the possible existence of a supersolid phase of a
two-dimensional dipolar crystal using quantum Monte Carlo methods at zero
temperature. Our results show that the commensurate solid is not a
supersolid in the thermodynamic limit. The presence of vacancies or
interstitials turns the solid into a supersolid phase even when a tiny
fraction of them are present in a macroscopic system. 
The effective interaction between vacancies
is repulsive making a quasiequilibrium dipolar supersolid possible.
\end{abstract}

\maketitle

Quantum systems with a dominant dipolar interaction have received permanent
interest from the recent achievement of a Bose-Einstein condensed state of
chromium atoms with a large dipole moment~\cite{griesmaier}. The anisotropy
of the dipole-dipole interaction leads to new exciting quantum phases that
have been observed for the first time in fully quantum
systems~\cite{lahaye}. The experimental confirmation of these predicted
phases and the collapses induced by the attractive part of the interaction
can be even better realized if the permanent dipole moment of the particles
becomes larger. A promising system for this goal is a  stable gas of
ultracold heteronuclear molecules~\cite{meerakker}.

If the quantum dipoles are confined in a two-dimensional (2D) plane, and
all the dipole moments are perpendicular to the plane, the interaction is
always repulsive and therefore the system is stable at any density. Under
such spatial and orientational restrictions one looses relevant features
that can emerge when the attractive collapse is approached but, on the
other side, the stability of the 2D geometry allows for the possible
observation of a gas-solid quantum phase transition at high
densities~\cite{grisha,buchler}. A 2D environment is currently devised in
the field of cold quantum gases by very anisotropic traps  where the
confinement in one direction is so tight that the transverse motion is
frozen to zero-point oscillations~\cite{gorlitz}. Another physical system
where this 2D setup is relevant is the one of indirect excitons composed by
electrons and holes physically separated using two coupled quantum
wells~\cite{rapaport,eisenstein}. If the distance between the electron and
hole layers is significantly smaller than the electron-electron and
hole-hole distances the resulting excitons can be modeled as composite
bosons with a dipole-dipole interaction~\cite{lozovik,palo}.

Recent quantum Monte Carlo calculations at zero~\cite{grisha} and finite
temperature~\cite{buchler} have shown that a 2D homogeneous phase of
dipoles experiments a gas-solid phase transition when the density
increases. The equation of state of this solid phase, which forms a
triangular lattice, as well as its main structure properties are already
reported in these previous works. However, relevant questions such as the
possible superfluid signal and/or condensate fraction of the
zero-temperature dipolar crystal were not addressed so far.  In fact, there
is at present a renewed interest in the search of supersolid phases where
off-diagonal long-range order and spatial solid order are simultaneously
present~\cite{balibar}. Supersolids appear as a new intriguing state of
matter that was predicted long-time ago and only recently observed in
torsional oscillator experiments with solid $^4$He~\cite{chan}.

Within the framework of Bose-Hubbard Hamiltonians supersolid phases of
dipolar lattice bosons have already been predicted. Danshita and S\'a de
Melo~\cite{danshita} identify exotic phases as checkerboard and striped
supersolid phases by including in the model Hamiltonian the attractive part
of the dipole-dipole interaction and Trefzger \textit{et
al.}~\cite{trefzeger} find a pair-supersolid phase in a bilayer
configuration. The emergence of supersolid states when dipolar bosons are
confined in two-dimensional optical lattices is probably favored by the
free tuning of the localization strength of the external lattice potential
included in the model Hamiltonian. A different concern is the possible
formation of a supersolid phase in a continuum system where a solid is
formed at high density without the presence of any external localization
potential. In this work, we present the first study of supersolidity in 2D
dipolar bosons at zero temperature using quantum Monte Carlo methods that
rely merely on the microscopic Hamiltonian.

The triangular crystal phase of dipolar bosons is studied by means of the
diffusion Monte Carlo (DMC) method that it is nowadays a standard tool for
achieving exact solutions of many-boson systems at zero temperature
relying on a stochastic approach~\cite{boronat}. The Hamiltonian describing
the 2D system of $N$ dipoles is
\begin{equation}
H= -\frac{\hbar^2}{2m} \sum_{i=1}^{N} {\bm \nabla}_i^2 + \sum_{i<j}^{N}
V(r_{ij}) \ ,
\label{hamilto}
\end{equation}
$m$ being the mass and $V(r)$ the dipole-dipole potential
\begin{equation}
V(r) = \frac{C_{\text dd}}{4 \pi} \, \frac{1}{r^3} \ .
\label{poten}
\end{equation}
The constant $C_{\text dd}$ depends on the nature of the dipole-dipole
interaction and increases proportionally to the square of the individual
dipole moment. As in previous work, we define  characteristic units of
length $r_0=m C_{\text dd}/(4 \pi \hbar^2)$ and energy ${\cal E}_0=
\hbar^2/(m r_0^2)$ in such a way that the properties of the system are
governed by a dimensionless density $nr_0^2$, with $n$ the particle
density. In order to fix the symmetry of the system and reduce the
statistical variance  DMC introduces a trial wave function $\Psi$ that is
used for importance sampling. In a previous work~\cite{grisha}, the crystal
phase of dipoles was studied using for $\Psi$ a non symmetric model
(Nosanow-Jastrow (NJ)) since the focus was the determination of the
equation of state and the phase transition point, issues in which implicit
symmetrization is much less relevant. Obviously, the NJ trial wave function
can not be used in the present study since our goal is the determination of
superfluid signals in the solid and that is only possible assuming particle
indistinguishability. To this end,  in the present work we use a symmetric
model
\begin{equation}
\Psi (\bm R) = \prod_{i<j}^{N} f(r_{ij}) \, \prod_{I=1}^{N_{\rm cr}}
\left[ \sum_{i=1}^{N} g(r_{Ii})  \right]  \
\label{trial}
\end{equation}
that was first introduced in the study of solid $^4$He at zero
temperature~\cite{cazorla}. In Eq. (\ref{trial}), ${\bm R} = \{ {\bm
r}_1,\ldots,{\bm r}_N \}$, $f(r)$ is a two-body Jastrow correlation factor
chosen as in Ref.~\cite{grisha}, $g(r_{Ii})= \exp [- \alpha ({\bm r}_i -
{\bm r}_I)^2]$, and $N_{\rm cr}$ is the number of lattice sites of the
triangular crystal structure. This model wave function (\ref{trial}) makes
compatible the spatial solid order and the symmetry under the interchange
of particles avoiding the numerically unworkable use of permanents on top
of the NJ wave function.

\begin{figure}[t]
\centerline{
\includegraphics[width=0.82\linewidth,angle=0]{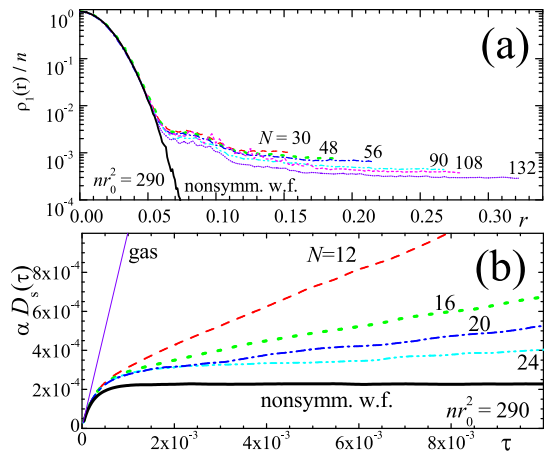}}%
\caption{(Color online) (a) One-body density matrix of the perfect 2D 
crystal at a density $nr_0^2=290$ and as a function of the number of 
particles $N$ used in the simulation; the solid line corresponds to a 
nonsymmetric trial wave function. (b) The function $\alpha D_{s}(\tau)$, 
Eq.~\ref{rhosuper}, at the same density and as a function of $N$; 
the thin straight line corresponds to $n_s/n = 1$ and the solid line to the 
nonsymmetric case, $n_s/n = 0$.}
\label{fig1}
\end{figure}

Coherence phenomena in the dipolar solid have been studied by calculating
the one-body density matrix $\rho_1(r)$ and the superfluid fraction
$n_s/n$.  The function $\rho_1(r)$ approaches a constant at long distances,
which is the condensate fraction $N_0/N=\lim_{r \rightarrow \infty}
\rho_1(r)/n$, if off-diagonal long range order exists in the system. In
DMC, the function  $\rho_1(r)$ can not be calculated using a pure estimator
and therefore some bias induced by the trial wave function $\Psi$ remains.
To reduce this bias as far as possible the variational parameters entering
in $\Psi$ have been optimized in such a way that the variational and DMC
(mixed) estimations of $N_0/N$ are coincident within their statistical
errors. On the other hand, the superfluid density is computed by extending
the winding-number technique, used in path integral Monte Carlo (PIMC)
simulations at finite temperature, to zero temperature~\cite{zhang}.
Explicitly,
\begin{equation}
\frac{n_{s}}{n} = \lim_{\tau \to \infty} \alpha
\left(\frac{D_{s} (\tau)}{\tau} \right) \ ,
\label{rhosuper}
\end{equation}
where $\alpha= N/ (4 D_0)$ with $D_0= \hbar^2/(2m) $, and $D_{s}
(\tau)=\langle({\bf R}_{\rm CM}(\tau) - {\bf R}_{\rm CM}(0))^2 \rangle$,
with $\bf{R}_{\rm CM}$ the center of mass of the particles and $\tau$ the
imaginary time. Differently from the estimation of $\rho_1(r)$, the measure
of the superfluid density (\ref{rhosuper}) is unbiased (pure estimator).

DMC results for the perfect 2D triangular solid are reported in
Fig.~\ref{fig1}.  In all the simulations, carried out with different number
of particles $N$, the one-body density matrix shows a plateau at long
distances and therefore a finite condensate fraction.  However, $N_0/N$
decreases significantly with $N$ making the condensate fraction vanishingly
small in the thermodynamic limit $N \rightarrow \infty$. If the calculation
is carried out with a nonsymmetric trial wave function (Nosanow-Jastrow
model), $\rho_1(r)$ does not show off-diagonal long range order for any
value of $N$ (see Fig.~\ref{fig1}). We show in the same figure results for
the superfluid density, plotting the function $\alpha D_{s} (\tau)$ (Eq.
\ref{rhosuper}) as a function of the imaginary time $\tau$; the slope of
this function is directly $n_{s}/ n$. As one can see, the slope becomes
zero within our numerical resolution for values $N \agt 30$ pointing to the
absence of supersolidity in the perfect crystal in the thermodynamic
limit.  The lack of supersolid signatures in the commensurate solid is
observed at any density, starting on the melting one $n_ {\text m}r_0^2 =
290(30)$ shown in Fig.~\ref{fig1}.

\begin{figure}[t]
\centerline{
        \includegraphics[width=0.82\linewidth,angle=0]{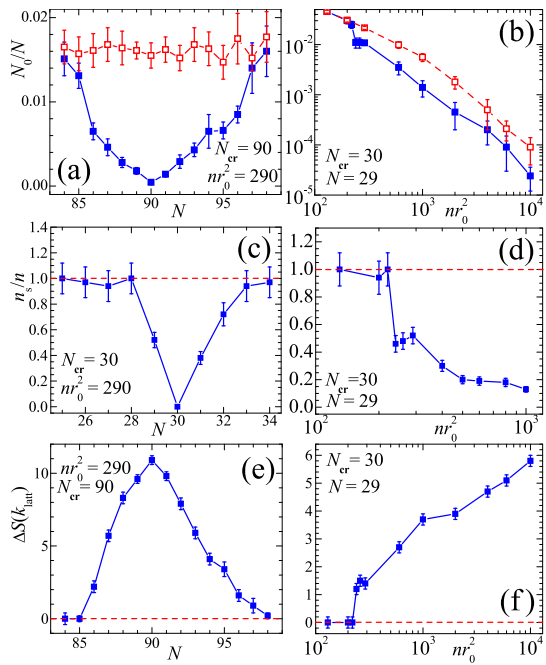}}%
        \caption{(Color online) Condensate fraction (a), superfluid
	fraction
	(c), and height of the main divergent peak of $\Delta S(k_{\rm latt})$ with 
	respect to the 	background at the reciprocal lattice
	vector $k_{\rm latt}$ (e), for a crystal with vacancies or interstitials ($N \neq
	N_{\rm cr}$). Density dependence of the condensate fraction (b),
	superfluid fraction (d), and $S(k)$ peak (f) for a solid with one
	vacancy. Solid and open points stand for the solid and gas phases,
	respectively.
	 }
\label{fig2}
\end{figure}

The presence of defects or imperfections in a crystal has been suggested as
a plausible explanation of the superolid signals observed experimentally in
torsional oscillator measurements of solid $^4$He. Whereas there are still
open discussions about the existence or not of vacancies in the
ground-state of bulk solid $^4$He~\cite{balibar}, it seems very reasonable
to think on the presence of vacancies or interstitials in a 2D crystal of
dipoles. Indeed, indirect excitons or trapped atoms or molecules with high
dipole moments can be easily produced with a fraction of defects. We have
collected in Fig.~\ref{fig2} DMC results for a dipolar solid with a finite
fraction of vacancies or interstitials. Both the condensate fraction and
superfluid density are very sensitive to the presence of defects: for any
concentration of vacancies or interstitials a finite value for $n_s/n$ and
$N_0/N$ is observed. When the fraction of vacancies is $\sim 6$\% and
$nr_0^2=290$, the supersolid melts: $N_0/N$ equals its value in the gas
phase, $n_s/n=1$, and the peak of $S(k)$ in the reciprocal lattice vector
disappears. The crystal also melts due to interstitials at a slightly
higher concentration, $\sim 10$\%.

In Fig.~\ref{fig2}, we show the density dependence of $N_0/N$, $n_s/n$, and
height of the main peak of $S(k)$ for the particular case of one vacancy in
a solid with $N_{\rm cr}=30$. The condensate fraction becomes vanishingly
small at high densities and remains always a factor of 3-5 smaller than its
value in the gas phase, as  also shown in the figure for comparison. The
superfluid density fraction is $\sim 50$\% at melting of the commensurate
crystal and decreases with $n$ but much more slowly than $N_0/N$. On the
other hand, the height of the $S(k)$ peak increases with density as
expected, and  it increases with $N$ for a fixed  $n$ as it must be in a
solid structure (not shown in the figure). At density $nr_0^2=230$ the
supersolid completely melts: $N_0/N$ becomes equal to  its value in the
gas, $n_s/n=1$, and the divergent peaks in $S(k)$ disappear.

As we commented before, the condensate fraction shows a significant
dependence with $N$ and therefore the estimation of the thermodynamic limit
when vacancies are present is fundamental. For this purpose, we performed a
study of the $N$-dependence of $N_0/N$ for vacancy fractions $0.018 \leq
N_{\text{vac}}/N_{\rm cr} \leq 0.042$. The DMC results obtained show a 
$1/N$ decrease with the number of Bose-condensed particles per vacancy,
$N_0/N_{\text{vac}}$, but with a finite value in the thermodynamic limit $N
\rightarrow \infty$ of $N_0/N_{\text{vac}}=0.050(8)$. The number of
superfluid particles  per vacancy, which is weakly dependent on $N$, also
remains finite in this limit.  Also, the height of the narrow peak in
$S(k)$ remains finite and proportional to $N$. Therefore, vacancy-induced
superfluidity coexists with spatial solid order, i.e., a supersolid phase
can exist.

\begin{figure}[t]
\centerline{
        \includegraphics[width=0.82\linewidth,angle=0]{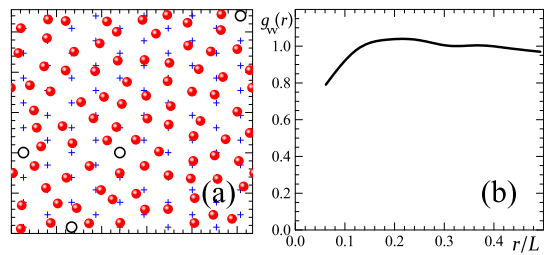}}%
        \caption{(Color online) (a) Snapshot of a typical spatial
	configuration when vacancies are present. This case corresponds to
	$N_{\rm cr}=90$ and $N=86$; crosses are the lattice points, solid circles
	are particles, and open circles, vacancies. (b) Vacancy-vacancy
	pair distribution function $g_{\text{vv}}(r)$ 
	for the same $N_{\rm cr}$ and $N$ values
	as in (a); $L=\sqrt{N/n}$ is the size of the simulation box.
	 }
\label{fig3}
\end{figure}

A relevant concern about the stability of a small fraction of vacancies in
the solid is the nature of their mutual interaction. Several microscopic
estimations in solid $^4$He show that two vacancies  tend to form a 
weakly-bound state because their interaction is
attractive~\cite{prokofev,ceperley,reatto}. Therefore, it has been
argued that vacancies would form a cluster inside the crystal that
eventually can evaporate producing a collapse of the crystal.
In order to characterize the local structure of vacancies in a 2D dipolar
crystal we have sampled the vacancy-vacancy two-body distribution function
$g_{\text{vv}}(r)$. As vacancies are not {\em real} particles and our
simulation works in a configuration space of particle coordinates one has
to define what a vacancy position is for a given snapshot of the system. In
our procedure, we have always identified a vacancy with one of the sites of
the perfect triangular lattice in which unambiguously none of the particles
is around it within a cutoff radius that is close to the value of the 
lattice constant. Along the evolution in imaginary time, there are
configurations in which we can not identify the vacancy sites due to
intrinsic fluctuations; in these cases we do not accumulate statistics for 
$g_{\text{vv}}(r)$.  In  Fig.~\ref{fig3}, we show a snapshot of the system
where vacancy sites are identified according to our definition. In the same
figure, the vacancy-vacancy correlation function $g_{\text{vv}}(r)$ is
shown for a setup composed by $N_{\rm cr}=90$ and four vacancies. The
radial function $g_{\text{vv}}(r)$ is normalized at each distance dividing
what accumulated in each bin by the corresponding output obtained in a
merely random distribution. As shown in Fig.~\ref{fig3}, vacancies repel at
short distances and this relevant feature is not only observed for this
particular set of parameters but settled at other densities and vacancy
fractions (below the threshold for melting). The repulsive interaction
between vacancies in the 2D dipolar solid is probably due to the
monotonously repulsive interaction between aligned dipoles, that makes 
configurations be more stable when vacancies spread in the system in  order
to effectively reduce the dipolar density. This is contrary to the
vacancy-vacancy attraction observed in solid $^4$He simulations in which
the van der Waals attraction at long distances can explain the difference
with the present results.

\begin{figure}[t]
\centerline{
        \includegraphics[width=0.82\linewidth,angle=0]{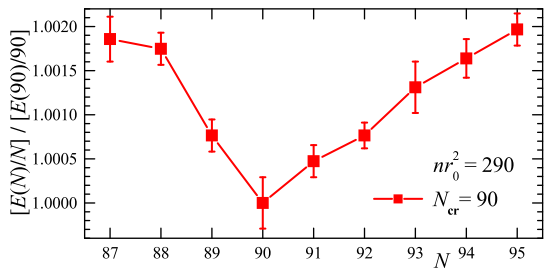}}%
        \caption{(Color online) Energy per particle of the 2D dipolar solid
	as a function of the number of vacancies or interstitials, at
	density $nr_0^2=290$ and $N_{\rm cr}=90$, and normalized to the energy per
	particle of the commensurate crystal.
	 }
\label{fig4}
\end{figure}

The ground state of the dipolar solid is a commensurate phase, i.e.,
without vacancies and/or interstitials. In other words, an activation
energy is needed to create a vacancy or interstitial. We have estimated the
activation energy to create one or more vacancies or interstitials using
the DMC method. In Fig.~\ref{fig4}, we show the energies per particle of a
solid with defects normalized to the energy per particle of the
commensurate solid and having rescaled the simulation box size to work at
fixed density. Our results support   the metastability of the solid with
defects with respect to the commensurate solid. The activation energy for
the creation of a vacancy is higher than the corresponding one for an
interstitial and this is maintained when the number of defects increases:
the slopes of the two cases are rather different (see Fig.~\ref{fig4}). It
is worth noticing that this behavior is opposite to the one observed in
solid $^4$He where the activation energy for an interstitial is
significantly larger than for a vacancy. Defining the activation energy in
the standard way~\cite{gillan}, we get for one vacancy
$E_\text{v}=2150(150)$ and for one interstitial $E_\text{i}=990(50)$, both
at fixed density  $nr_0^2=290$. These activation energies are significantly
larger than the Berezinskii-Kosterlitz-Thouless temperature
$T=380$~\cite{buchler},  making thermal activation of defects in a dipolar
crystal not possible.

Summarizing, we have studied the possible emergence of bosonic coherence
phenomena in a two-dimensional crystal of dipoles by calculating the
condensate fraction and superfluid density using accurate quantum Monte
Carlo methods. To this end, we have used for the first time in this system
a trial wave function for importance sampling with both boson symmetry and
solid order. Our DMC results show that the commensurate solid is not a
supersolid since both $n_s/n$ and $N_0/N$ become zero  in the thermodynamic
limit within our numerical resolution. The introduction of defects, in the
form of vacancies or interstitials, produces a dramatic effect on both
quantities, even with tiny concentrations. A quasiequilibrium solid with
vacancies or interstitials is proven to be supersolid within a predicted
fraction of defects. If this percentage is further increased the supersolid
melts. The effective vacancy-vacancy interaction is repulsive at short
distances, a feature that is opposite to the one of solid $^4$He and that
can help to stabilize the dipolar supersolid phase. The recrystallization
to the ground state is exponentially suppressed by the tunelling barrier
which stabilizes a dipolar supersolid with a small fraction of defects.
Possible experimental realizations of a quasiequilibrium dipolar supersolid
with defects include i) harmonically trapped dipolar
molecules~\cite{damski} or atoms~\cite{goral} spatially localized with an 
optical lattice and ii) Wannier-Mott 2D dipolar excitons in single or
coupled semiconductor quantum wells in electric and magnetic fields which
are perpendicular to the quantum wells plane~\cite{filinov}. In the latter
case, the finite excitation lifetime caused by their optical recombination
gives rise to a continuous addition of vacancies into the system, resulting
in a macroscopic supersolid at low temperatures.

We wish to thank partial financial support from DGI (Spain) Grant No.
 FIS2008-04403, Ram\'on y Cajal Program, and Generalitat de Catalunya Grant No. 2005SGR-00779 is also
 akcnowledged.

\end{document}